\documentclass[]{spie}  
\usepackage[]{graphicx}
\usepackage{booktabs}
\usepackage{subfigure}

\title{Conceptual Design of the Coronagraphic High Angular Resolution Imaging Spectrograph (CHARIS) for the Subaru Telescope} 
\author{Mary Anne Peters\supit{a}, Tyler Groff\supit{a}, N. Jeremy Kasdin\supit{a}, Michael W. McElwain\supit{b}, Michael Galvin\supit{a}, Michael A. Carr\supit{a}, Robert Lupton\supit{a}, James E. Gunn\supit{a}, Gillian Knapp\supit{a}, Qian Gong\supit{b}, Alexis Carlotti\supit{a}, Timothy Brandt\supit{a}, Markus Janson\supit{a}, Olivier Guyon\supit{c}, Frantz Martinache\supit{c}, Masahiko Hayashi\supit{c}, Naruhisa Takato\supit{c}
\skiplinehalf
\supit{a}Princeton University, Princeton, NJ, USA; \\
\supit{b}Goddard Space Flight Center, Greenbelt, MD, USA; \\
\supit{c}Subaru Telescope, National Astronomical Observatory of Japan, Hilo, HI, USA;\\
}

\authorinfo{Further author information: (Send correspondence to Mary Anne Peters)\\E-mail: mapeters@princeton.edu, Telephone: 1 609 258 0096}
 
\begin{document} 
\maketitle 

%%%%%%%%%%%%%%%%%%%%%%%%%%%%%%%%%%%%%%%%%%%%%%%%%%%%%%%%%%%%% 
\begin{abstract}
Recent developments in high-contrast imaging techniques now make possible both imaging and spectroscopy of planets around nearby stars. We present the conceptual design of the Coronagraphic High Angular Resolution Imaging Spectrograph (CHARIS), a lenslet-based, cryogenic integral field spectrograph (IFS) for imaging exoplanets on the Subaru telescope. The IFS will provide spectral information for 140$\times$140 spatial elements over a 1.75 arcsecs $\times$ 1.75 arcsecs field of view (FOV). CHARIS will operate in the near infrared ($\lambda = 0.9 - 2.5 \mu m$) and provide a spectral resolution of R = 14, 33, and 65 in three separate observing modes. Taking advantage of the adaptive optics systems and advanced coronagraphs (AO188 and SCExAO) on the Subaru telescope, CHARIS will provide sufficient contrast to obtain spectra of young self-luminous Jupiter-mass exoplanets. CHARIS is in the early design phases and is projected to have first light by the end of 2015. We report here on the current conceptual design of CHARIS and the design challenges. 
\end{abstract}


%%%%%%%%%%%%%%%%%%%%%%%%%%%%%%%%%%%%%%%%%%%%%%%%%%%%%%%%%%%%%
\section{INTRODUCTION}
\label{sec:intro} 
Over 700 exoplanets have been confirmed via indirect methods over the past seventeen years\footnote{exoplanets.org}. However, Doppler and transit techniques only probe the exoplanet's orbital parameters and are sensitive to short-period exoplanets.  Direct imaging coupled with spectroscopy represents one of the next big steps in the exoplanet field. It is sensitive to exoplanets at large separations, complementing the Doppler and transit techniques and it allows for spectral characterization of exoplanets. The exoplanet community has recently begun to image exoplanets \cite{Marois2008HR8799,Lagrange2008A-probable}. The addition of spectroscopy to direct imaging will be particularly useful in characterizing exoplanets that are like Earth and may support life \cite{Kawahara2012Can-Ground-based}.  Integral field spectrographs (IFSs) are well suited for taking spectra of exoplanets.  

An IFS simultaneously obtains spatial and spectral information over the field of view by dispersing the entire image on the detector. Our design uses a lenslet-based IFS to sample the image plane. Each lenslet samples a piece of the image and focuses it to a point spread function (PSF). Each PSF is dispersed and then imaged by the detector. This allows the IFS to measure two spatial and one spectral dimension simultaneously. This type of lenslet array-based spectrograph was first implemented on the TIGER IFS\cite{bacon19953D} at visible wavelengths. Several mid- to high- resolution IFSs have since been built and are on-sky including OSIRIS\cite{larkin2006osiris} and GFP-IFS\cite{Bonfield2008GFP}. CHARIS is a low resolution IFS similar to GPI\cite{macintosh2008gemini, Perrin2010The-Integral}, SPHERE\cite{beuzit2008sphere, Claudi2011Optical} and Project 1640\cite{Hinkley2010A-New-High} in that it is designed to image exoplanets. Combined with the SCExAO and AO188 systems, CHARIS will be the first exoplanet-purposed IFS on an 8m class telescope in the northern hemisphere able to achieve a small inner-working angle ($2 \lambda/D$) and contrasts of  $ 10^{-4}-10^{-7} $\cite{McElwain2012Scientific}. It is also unique in that it will offer both a low-resolution mode (R=14) able to simultaneous collect photons from $\lambda=0.9-2.4 \mu m$ (y, J, H, and K bands) and a high spectral resolution, R=65 over a single bandpass (y, J, H or K band) and a 1.75 arcsecs $\times$ 1.75 arcsecs field of view (FOV). 

CHARIS is currently in the early design phases, and is being built at Princeton University under PI Jeremy Kasdin. We completed the CHARIS conceptual design review in March 2012, and the preliminary design review is scheduled for December 2012. Fabrication will begin in summer 2013, and CHARIS is scheduled to be on-sky by the end of 2015. In this paper we discuss the optical design (section \ref{sec:optical}) and innovative methods to minimize crosstalk between spatial elements (section \ref{sim}). Section \ref{sec:mech} details the conceptual mechanical design of the CHARIS instrument.  The CHARIS scientific design is discussed in McElwain et al. (2012)\cite{McElwain2012Scientific}.

%%%%%%%%%%%%%%%%%%%%%%%%%%%%%%%%%%%%%%%%%%%%%%%%%%%%%%%%%%%%%
\section{OPTICAL DESIGN} \label{sec:optical}

\subsection{High Contrast Instrumentation at Subaru}

\begin{figure}
\begin{center}
\begin{tabular}{c}
\includegraphics[scale = 0.6]{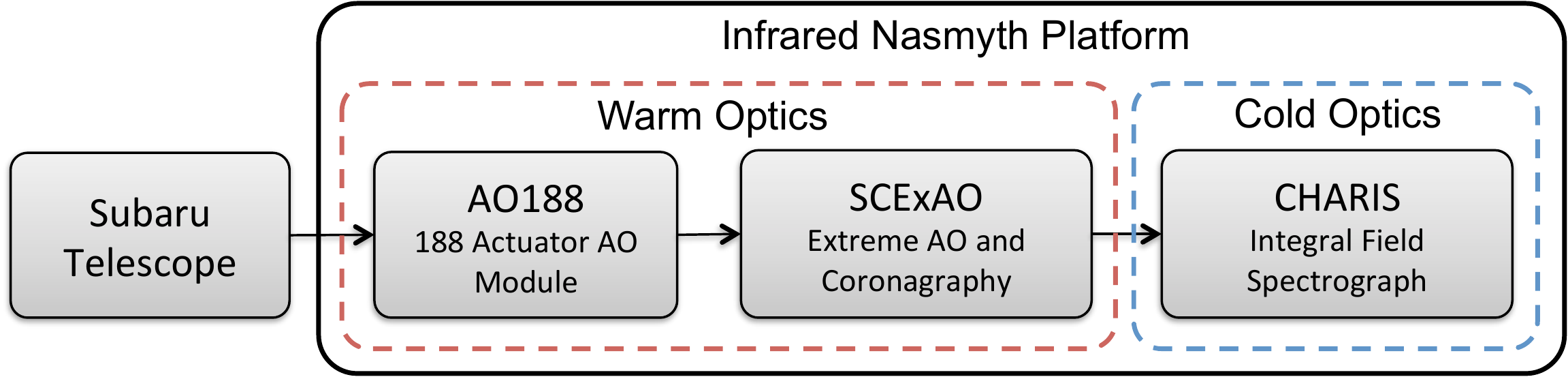}
\end{tabular}
\end{center}
\caption[example] 
{ \label{fig:block} 
Block diagram of the instruments prior to and including the CHARIS IFS. The Subaru telescope focuses the light into AO188 (the primary AO system) on the infrared Nasmyth platform. The beam is then sent into SCExAO which has an extreme AO system and coronagraphic capabilities.}
%\vspace{7 mm}
\end{figure} 

The CHARIS instrument will be located on the infrared Nasmyth platform of the Subaru telescope. It will be preceded by AO188 and SCExAO (see Figure \ref{fig:block}). CHARIS will be interchangeable with the current high-contrast imaging camera (HiCIAO)\cite{Hodapp2008HiCIAO}. HiCIAO, in conjunction with the SEEDS program, has discovered cold substellar/planet companions to several nearby stars \cite{Thalmann2009Discovery} and fine structure, indicative of planet formation, in many protoplanetary disks \cite{Dong2012The-missing, Hashimoto2011Direct, Muto2012Discovery}. 

HiCIAO currently operates behind the AO188 adaptive optics (AO) system. The SCExAO extreme AO system is currently in the commissioning phase. Once commissioning is complete, HiCIAO will be used as the science camera for AO188 or AO188+SCExAO. AO188 is equipped with a 188-element wavefront curvature sensor with photon counting APD modules and a 188 element bimorph mirror \cite{Hayano2010Commissioning, Hayano2008Current}. In typical seeing conditions on Mauna Kea in Hawaii, AO188 achieves K-band Strehl ratios of 60\% to 70\% with a 9th magnitude (R) natural guide star at a 30 arcsecs isoplanatic angle. The AO system works with guide stars down to R = 16.5 \cite{minowa2010performance}.

The SCExAO instrument includes advanced coronagraphic techniques (such as Phase Induced Amplitude Apodisation or PIAA) and a high order, extreme AO system designed specifically for high contrast imaging of exoplanets and circumstellar disks. When CHARIS is on-sky, SCExAO will be capable of producing H-band Strehl ratios of $\sim$90\%. The instrument uses two infrared wavefront sensors (WFS) and a fast visible WFS to drive a single MEMS  deformable mirror that is on a tip-tilt mount. The WFSs and control architecture are integrated with the coronagraph system. More information on the SCExAO system is given by Guyon et al. (2011) and Martinache et al. (2011)\cite{guyon2011wavefront, martinache2011subaru}. 

\subsection{Optical Design of CHARIS}

CHARIS will use several methods to obtain high contrast imagery. It is a goal to implement spectral focal plane wavefront control using the CHARIS science imagery in real time in conjunction with the SCExAO system. SCExAO and CHARIS will also implement advanced coronagraphy techniques including the PIAA coronagraph\cite{Guyon2005Exoplanets} and shaped pupils\cite{kasdin2003extrasolar}. CHARIS may also include the ability to do non-redundant masking (NRM), which could aid in the detection of exoplanets\cite{Kraus2011LkCa}.

\begin{figure}
\begin{center}
\begin{tabular}{c}
\includegraphics[scale = 0.83]{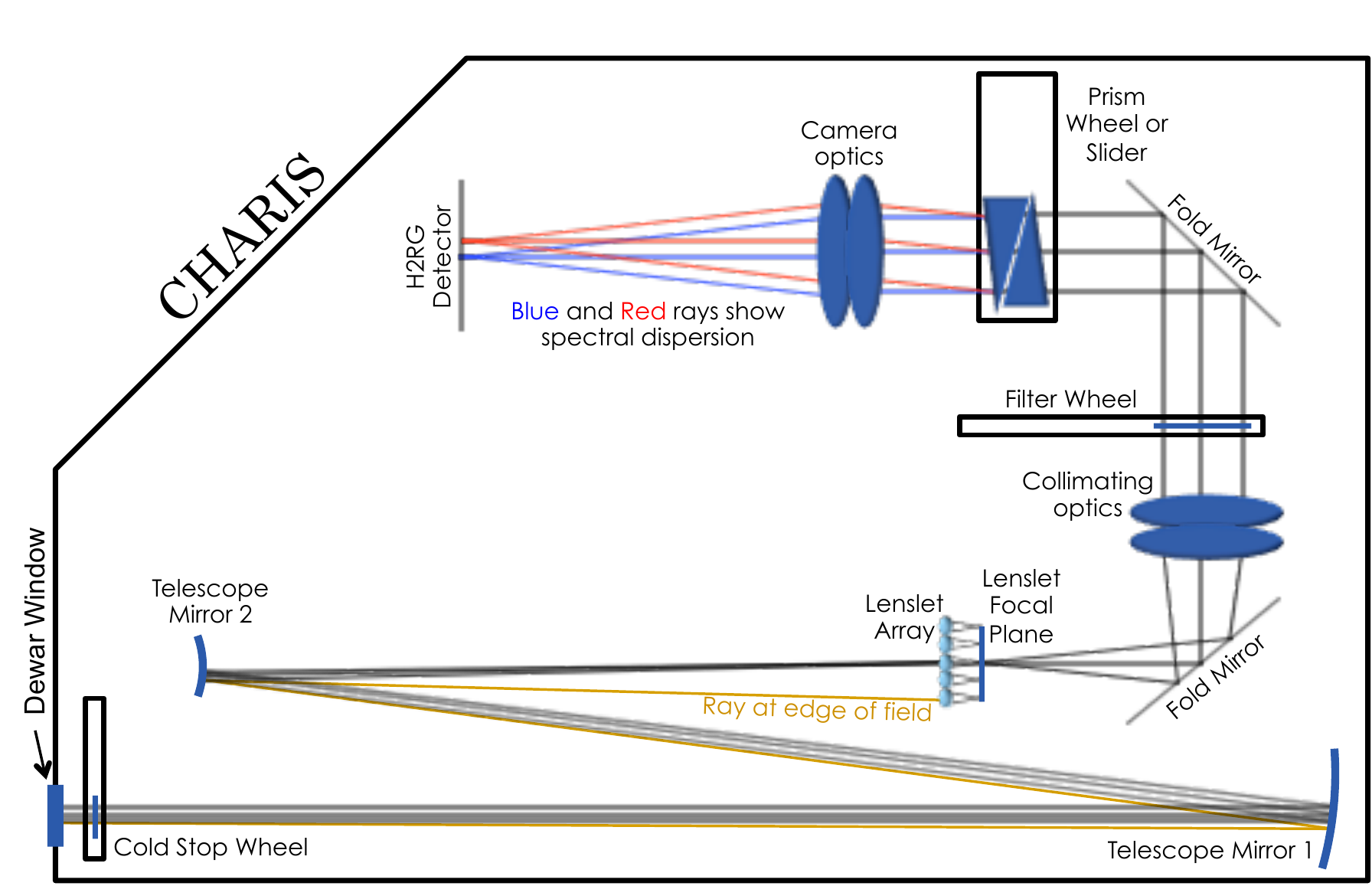}
\end{tabular}
\end{center}
\caption[example] 
{ \label{fig:optical-layout} 
Optical layout of the CHARIS instrument (roughly to scale).}
\end{figure} 

\begin{figure}
\begin{center}
\begin{tabular}{c}
\includegraphics[scale = .85]{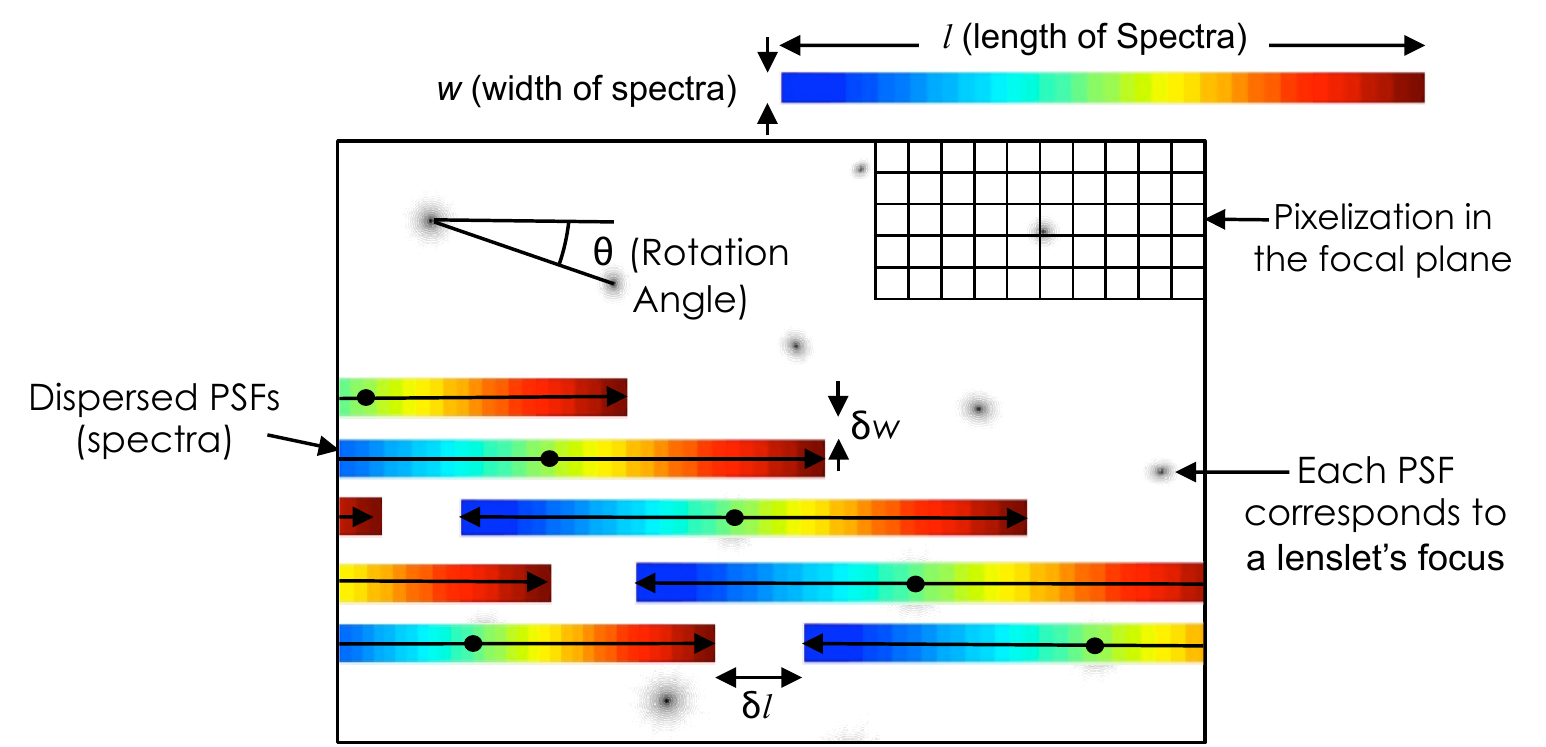}
\end{tabular}
\end{center}
\caption[example] 
{ \label{rotation} 
The layout of the spectra and pre-dispersed PSFs on the detector. The rotation of the dispersion relative to the x-axis of the detector ($\theta$) allows for optimal use of the pixels on the detector and increases the available area ($wl$) for dispersion. Shown here are $l$, the length of the spectra, $\delta l$, the horizontal gap between the spectra, $w$ the width of the spectra, and $\delta w$ the vertical gap between the spectra. The pixelization shown on the upper right is to scale relative to the length of the spectra and size of the PSFs. The process to create the PSFs (shown here as small black dots) is shown in Figure \ref{fig:Lenslets}.}
\end{figure}

A sketch of the functional components in CHARIS is shown in Figure \ref{fig:optical-layout}. CHARIS receives a collimated beam $\sim$10mm in diameter from SCExAO. The light  propagates through a cold stop in the pupil plane. The cold stop is placed in a wheel and can be interchanged with a non-redundant mask or a shaped pupil coronagraph. PIAA is in SCExAO rather than CHARIS because it requires two pupil planes. The beam is then focused onto the lenslet array using Cassegrain optics to obtain a slow beam (F/500). The lenslet array spatially samples the light and focuses the incident light on each lenslet into a PSF at the lenslet focal plane (see Figure \ref{rotation}). The light is then collimated and dispersed by a prism pair. We rotate the lenslet array or the prism pair to disperse the spectra at an angle to separate the spectra and avoid overlap on the detector (see Figure \ref{rotation}). The light is then filtered to the appropriate bandpass prior to dispersion and focused onto the Hawaii-2RG detector.

The current conceptual design for the IFS includes a baseline spectral resolution of R = 33 with a bandpass of $\Delta\lambda = 0.7 \mu m$ (J+H band or H+K band). It is a goal to include an R = 14 mode with a $\Delta\lambda = 1.4 \mu m$ bandpass (y, J, H and K band can be measured simultaneously) and an R = 65 mode with a $\Delta\lambda = 0.4 \mu m$ bandpass. The spectral resolution will be changed by switching dispersive elements and changing the bandpass filter. All three configurations contain 140$\times$140 spatial measurements (or spaxels) and 16 spectral measurements. For all three modes, the plate scale at the lenslet array will be 12.6 milli-arcsecs (mas) such that the shortest wavelength of $\lambda = 0.9\mu m$ is Nyquist sampled. 

The current control radius of the SCExAO DM is $16\lambda/D$, although a larger DM, expected before CHARIS is on-sky, should allow for control out to $\sim 32\lambda/D$. The field of view was chosen such that the $16\lambda/D$ control radius is accommodated for all wavelengths, and the $ 32\lambda/D$ control radius is accommodated in z-band and J-band. The distances to nearby stars and planet formation theories suggest that most discoveries will occur within 2 arcsecs separation from the host star. The key design parameters for the CHARIS IFS are listed below.

\begin{center}
\begin{tabular}{ llrrr }
    $D$ & Diameter of Telescope & 8.2$m$\\
    $q$ & Detector Pixel Pitch & 18$\mu m$\\         
    $N_{Pixels}$ & Number of Pixels on the Detector & 2048 pixels $\times$ 2048 pixels\\ 
    $\lambda_{max}$ & Maximum Wavelength & 2.4$\mu m$  \\
    $\lambda_{min}$ & Minimum Wavelength & 0.9$\mu m$\\ 
    $\theta$ & Plate Scale at the Lenslet Array & 12.6$mas$\\
    $\Delta\lambda$ & Bandpass & 0.7$\mu m$$^a$ & 1.5$\mu m$$^b$ & 0.4$\mu m$$^c$\\
    $R$ & Spectral Resolution & 33$^a$ & 14$^b$ & 65$^c$\\
    $N_{spectral}$ & Number of spectral measurements & 16\\
    $N_{spatial}$ & Number of spatial measurements & 140$\times$140 \\ 
    $FOV$ & Field of View & 1.75 $arcsecs$ $\times$ 1.75 $arcsecs$\\
    $l+\delta l$ & Length of a spectrum + gap & 35 pixels\\
    $w+\delta w$ & Width of spectrum + gap & 6 pixels\\
\end{tabular}

$^a$Baseline design;
$^b$Low Resolution design;
$^c$High Resolution design
 \label{table2}
\end{center}
% \end{table}

\subsection{Lenslet Array Design} \label{sim}
The lenslet (or microlens) array is the pivotal optical element of an IFS. This optic influences the spectral separation and the layout of the spectra on the detector as shown in Figure \ref{rotation}. Crosstalk, or light from adjacent lenslets or spaxels overlapping on the focal plane, can be a major source of noise in lenslet-based IFSs. We are investigating new ways to eliminate this crosstalk noise. We have developed a MATLAB model to simulate the lenslet array, estimate the crosstalk noise and simulate CHARIS imagery. The basic MATLAB model allows the user to vary the shape (square or circle), diameter and f-number of the lenslets, FOV, detector pitch, rotation angle, and the f-number of the light incident on the lenslet array. We have included a pupil plane mask (coronagraph or NRM), a Kolmogorov phase field (to vary Strehl), and an off-axis source to simulate an exoplanet. The initial input field for the code is the pupil of the telescope. The model uses Fresnel approximation propagation. The model was used to investigate various IFS parameters in order to determine the optimal design.

We have begun to validate the model both analytically and with a separate Zemax model. For the simple case of a circular on-axis lenslet, the Zemax physical optics propagation tool predicts the same encircled energy to within 5\% of the MATLAB model. When we input the design for the SPHERE IFS, the MATLAB model reproduces the SPHERE lenslet PSF within $\sim10\%$ of what their team measured in the lab \cite{Claudi2011Optical}. In addition, a lab bench with lenslet array and imaging optics is currently being setup at Princeton to verify the accuracy of the model. 

\begin{figure}
\begin{center}
\begin{tabular}{c}
\includegraphics[scale = 0.37]{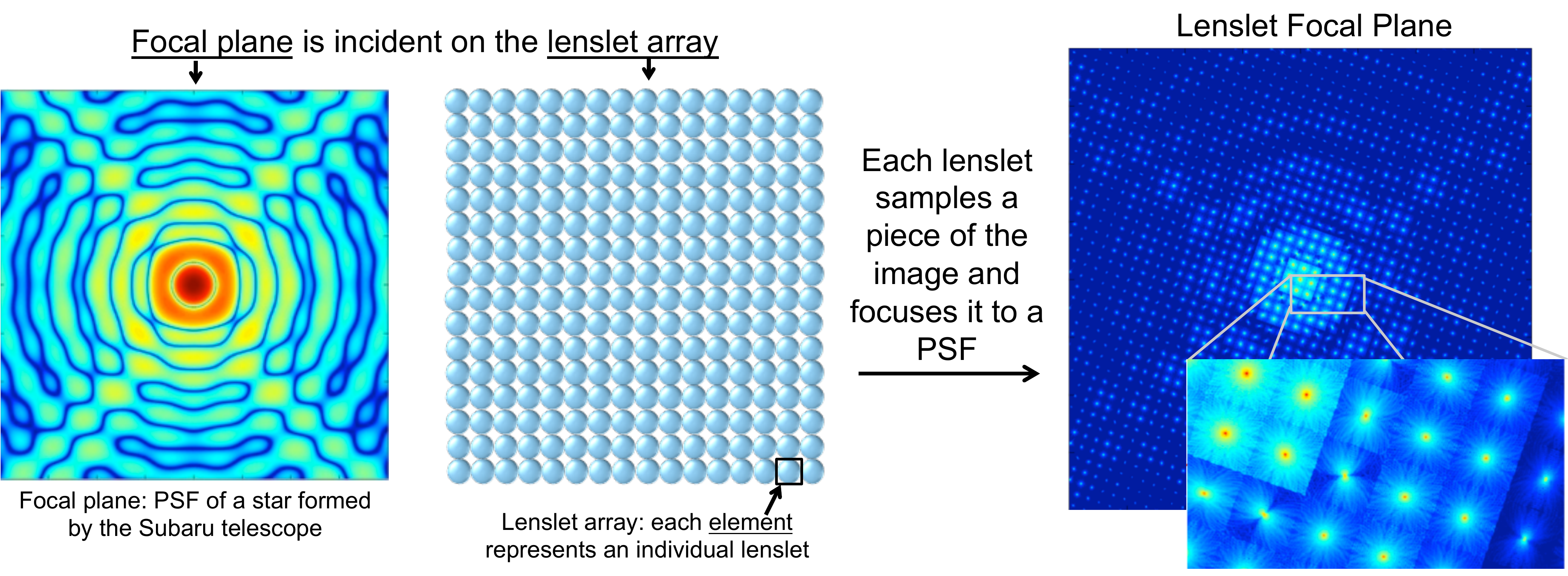}
\end{tabular}
\end{center}
\caption[example] 
{ \label{fig:Lenslets} 
This figure demonstrates the lenslet portion of the MATLAB model. On the left is the PSF of the Subaru telescope (labeled ``Focal Plane"). The focal plane is sampled by the lenslet array (circular lenslets are used in this simulation). Each lenslet focuses the light into small spots shown on the far right (labeled ``Lenslet Focal Plane"). A magnified section of the lenslet focal plane is shown to demonstrate the structure of each individual PSF. The diffraction ringing outside the core of the PSF (shown on the right) leads to crosstalk. All images are plotted on a log scale with a $10^8$ dynamic range. Circular lenslets are used in this simulation. Note that the PSFs on the right correspond with the PSFs shown as black points or PSFs in Figure \ref{rotation} before the spectral dispersion.}
\end{figure} 

Figure \ref{fig:Lenslets} illustrates the lenslet modeling portion of the MATLAB model. For the case shown here, we assume no atmosphere, coronagraph, or exoplanet. The model Fourier transforms the Subaru pupil into a focal plane (top left of Figure \ref{fig:Lenslets}). The small portion of the focal plane incident on a single lenslet is then Fourier transformed by that lenslet and forms a  PSF at the lenslet's focus. We refer to this as the lenslet focal plane (top right of Figure \ref{fig:Lenslets}) because this is where the focus of the lenslets occur.  If the dispersive element is removed from the system or if we are looking at only one wavelength of light, the lenslet focal plane is a good representation of the image on the detector (ignoring any magnification). In the MATLAB model, we use the image in the lenslet focal plane to calculate the energy encircled within a 2x2 pixel area at the center of each spaxel and the cross-talk that would occur if each spot were dispersed into a spectrum. Based on this model we chose F/5, $250 \mu m$ diameter lenslets, which focus more than 95\% of the energy into the central 2x2 pixels at all wavelengths (0.9-2.5 microns). We found that slow incident f-numbers and fast lenslet f-numbers led to the higher energy encircled values. We chose square lenslets for their high fill factor; a typical fill factor for square lenslets is $\sim$96\% compared to $\sim$75\% for circular lenslets. A full list of the lenslet parameters is given below.

%\begin{table}
%\caption{CHARIS Lenslet Parameters}
\begin{center}
\begin{tabular}{ lrclr }
    Diameter of the Lenslets & 250 $\mu m$ && Energy Encircled in 2x2 pixel area & $>96\%$\\ 
    F/\# of the lenslets & F/5 && Magnification from Lenslet to Detector & 1.23\\
    F/\# incident on lenslet array & F/500 && Lenslet Rotation Angle & $26.5^{\circ}$ \\        
    Lenslet Shape & Square && Perpendicular Gap between Spectra & 4.2 pixels\\ 
    Lenslet Packing & Square && Parallel Gap between Spectra & 3.0 pixels\\
\end{tabular}
 \label{table2}
  \end{center}
 %\end{table}

While these lenslets focus $\sim 95\%$ of the light into a 2x2 pixel area, the remaining $\sim 5\%$ of the light can still cause cross-talk  with adjacent spectra on the detector.  The crosstalk is particularly problematic when a coronagraph produces high contrasts between adjacent spatial locations in the field, which the science camera should preserve. For the coronagraphs that are likely to be used with CHARIS (shaped pupils and the PIAA), a crosstalk level of $<10^{-4}$ is acceptable  between adjacent spatial locations. This value is derived from the maximum contrast that is achievable between any two adjacent lenstlets. The remaining light needs to be suppressed or eliminated. We are currently exploring several methods to suppress the crosstalk. One possible solution would be to use a pinhole array in the focal plane after the lenslet array to block the unwanted light (discussed further in Section \ref{pinholearray}). Another method would be to apodize the lensletÕs entrance aperture such that the ringing from the diffraction is redirected outside of the lenslet array (discussed further in Section \ref{lensletapodization}). Both methods are, in theory, capable of reducing the crosstalk to $\sim10^{-4}$ from the nominal $\sim10^{-1}$-$\sim10^{-2}$ crosstalk noise levels that typically occurs with no suppression techniques.

\begin{figure}
\begin{center}
\begin{tabular}{c}
\includegraphics[scale = 0.8]{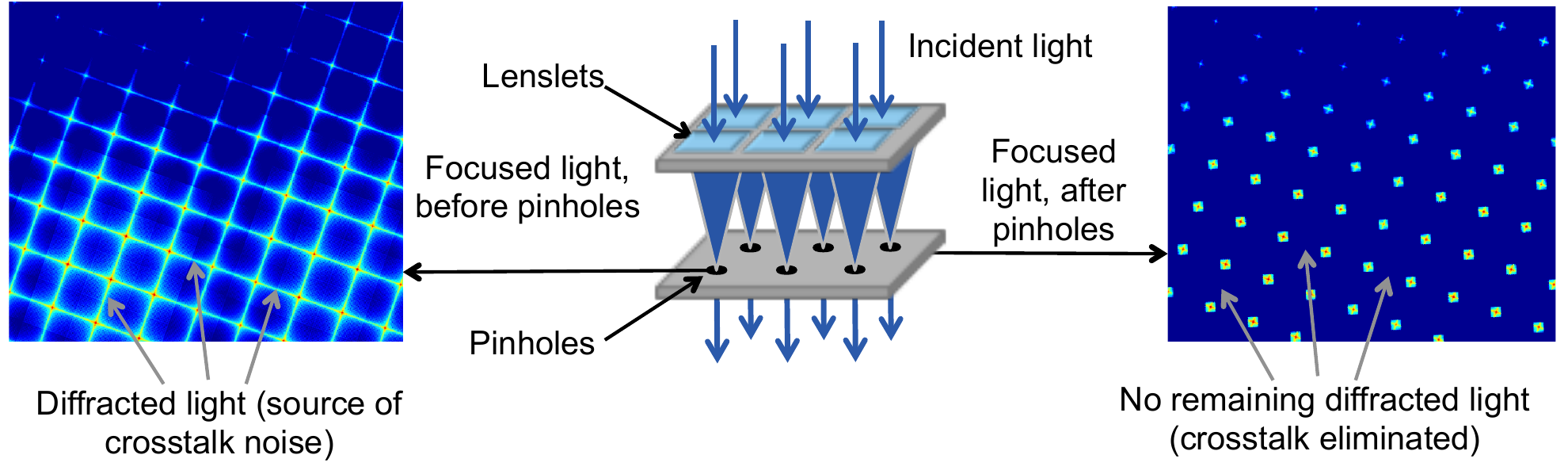}
\end{tabular}
\end{center}
\caption[example] 
{ \label{pinholes} 
Diffraction spots at the focus of the lenslets before (left) and after (right) a pinhole array. The pinholes block the diffracted light, decreasing crosstalk noise. Center - sketch of the lenslets and pinhole array with the path of the light shown in dark blue arrows and cones. Square lenslets were used in this example.}
\end{figure} 

\subsubsection{Pinhole Array} \label{pinholearray}

A pinhole array could be placed in the focal plane after the lenslet array to block the diffraction rings or spikes (see Figure \ref{pinholes}). This is similar to the pinhole array described by Bonfield et al. (2008)\cite{Bonfield2008GFP} and implemented in the GIII instrument at the Apache Point Observatory. The advantage of this solution is that, by blocking all light outside of a small opening, pinholes could achieve $10^{-4}$ contrast. Unfortunately, phase errors and phase tilts due to optics and atmospheric turbulence will displace the lenslet's focus from the center of the pinhole. As an aside, this offset is the same principle that is used to measure phase angle in a Shack-Hartmann wavefront sensor. If the phase error is too large, the pinhole will vignette some or all of the light. Furthermore, if the phase error changes with time or spatial location on the lenslet array, the amount of light lost in each pinhole will be difficult to calibrate. If pinholes are used in CHARIS, the phase error, which could result from atmospheric turbulence, optical phase aberrations, and a non-telecentric beam incident on the lenslet array, must be tightly constrained. The MATLAB model described in Section \ref{sim} has the capability to simulate these effects. If pinholes are used to eliminate crosstalk in CHARIS, the MATLAB simulations will simulate atmospheric turbulence, non-telecentricity and optical phase errors, and determine the maximum allowable phase error at the lenslet array. The lab bench lenslet setup that is being used to test the MATLAB model will allow us to verify this method and check the accuracy of the simulated cross-talk suppression.

\subsubsection{Lenslet Apodization} \label{lensletapodization}

Shaped or apodized pupils, such as those discussed by Kasdin et al. (2003)\cite{kasdin2003extrasolar}, have been proposed as a coronagraphic technique for imaging exoplanets. Shaped pupils are a binary mask that redistribute light in the focal plane by blocking light in the pupil plane in such a way that the Fourier transform of that shape produces the desired shape. Similar to shaped pupils, apodized pupils redistribute the light in the focal plane, however the apodized pupil masks are not binary, but consist of continuous apodization or grayscales. In this case, we want a shaped or apodized pupil to place the diffracted light outside the lenslet array altogether to eliminate the crosstalk. To implement this method, we would procure a lenslet array with a flat front surface (no optical power) and a powered back surface to focus the light. This would allow us to place the shaped or apodized pupil on the front surface of the lenslet array. Apodizing the lensletÕs entrance aperture reduces crosstalk at the expense of throughput. This is a tradeoff that must be carefully considered if this method is used. 

The primary purpose of the shaped or apodized pupil is to decrease crosstalk, however another advantage is that it broadens the lenslet's PSF and therefore distributes the light more uniformly across the 2x2 pixel area sampling the PSF. Unlike in the pinhole array the shaped pupil will not block or vignette the light if there are phase errors prior to the lenslet array. Phase errors can still lead to a slight shift of the PSFs and spectra on the detector. However, the gap between spectra ($\delta l$ and $\delta w$) can be chosen such that the movement of the spectra due to phase errors is less than $\delta l$ and $\delta w$. Apodization has the advantage of eliminating the pinholes which avoids vignetting the beam. 

Because of the small size of the lenslets, we have a limited number of points and that may not be enough to make an adequate shaped pupil.  Approximately 250$\times$250 points would be possible on each lenslet, corresponding to 1 $\mu m$ features. The coarse number of spatial elements and the large outer working angles (diffracted light needs to be sent outside the lenslet array) makes it near impossible to design a binary shaped pupil mask that can achieve $10^{-4}$ contrast. It may be that for the contrast we are trying to achieve, a commercial apodization of the lenslets will be possible. The continuously apodized pupils have a much larger outer working angle than shaped pupils, and therefore it may be possible to design such an apodized pupil that meets our requirements. However, the combination of a pinhole mask and apodized pupil could provide the simplest solution, which we describe in the following section.

\begin{figure}
\begin{center}
\begin{tabular}{c}
\includegraphics[scale = 0.78]{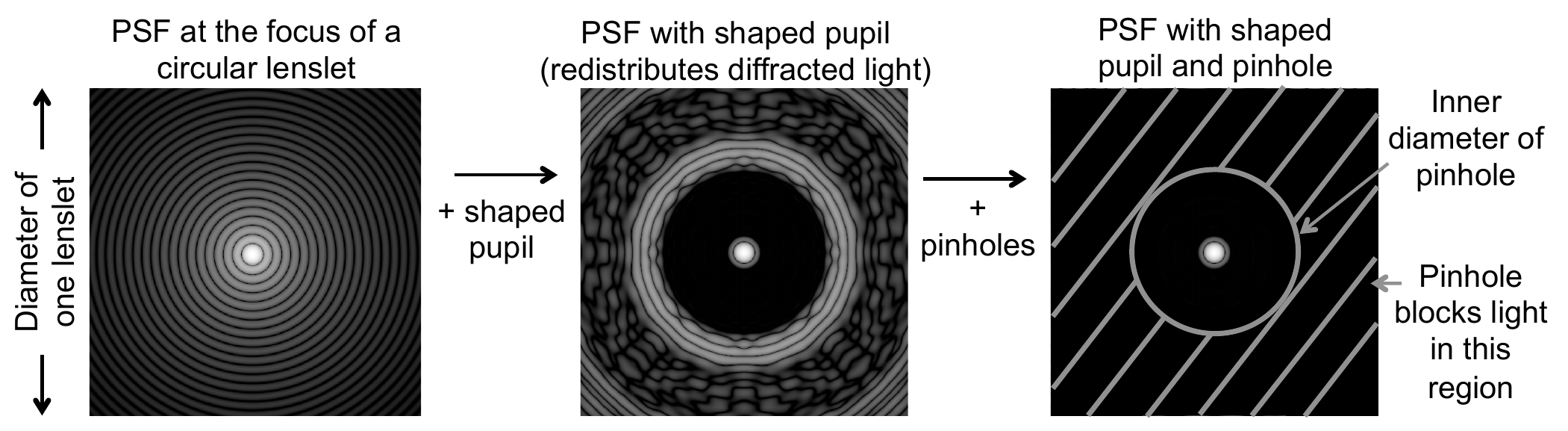}
\end{tabular}
\end{center}
\caption[example] 
{ \label{shapedpupils}
Left - PSF (airy pattern) at the focus of a circular lenslet. The shaped pupil separates the diffracted light from the core of the airy pattern (center). This allows for the unwanted diffracted light that would lead to crosstalk to be blocked by a pinhole (right). The combined pinholes + shaped pupil concept may reduce crosstalk noise to $<10^{-4}$. Pinhole allows for a smaller outer working angle, and the shaped pupil allows for a larger pinhole diameter.}
\end{figure} 

\subsubsection{Shaped Pupils + Pinholes}

By combining the apodized or shaped pupil and pinhole methods, we may relax the design constraints associated with each. For this case we can reduces outer working angle constraint of apodized pupil because the pinholes can be used to block the diffracted light.  The apodization in turn relaxes the tolerance of the pinhole diameter, thus reducing phase error requirements. The idea would be to use pinholes that are approximately half the diameter of the lenslets and use the apodized pupils to place the diffracted light in the region where it would be blocked by the pinholes. The larger pinholes relax the phase error requirements, while the apodized pupil now only needs to redirect the diffracted light by half the diameter of a lenslet. This decrease in outer working angle may even allow for shaped pupils to be used rather than apodization.  Figure \ref{shapedpupils} illustrates our proposed design.  We are currently optimizing apodized and shaped pupils that meet our requirements. We are capable of fabricating shaped pupils at Princeton, and will be able to test the shaped pupil plus pinhole design this summer. Further modeling and optical bench tests will verify if the crosstalk is sufficiently suppressed using this method.

%%%%%%%%%%%%%%%%%%%%%%%%%%%%%%%%%%%%%%%%%%%%%%%%%%%%%%%%%%%%%
\section{MECHANICAL DESIGN} \label{sec:mech}

\begin{figure}
\begin{center}
\begin{tabular}{c}
\includegraphics[scale = 0.73]{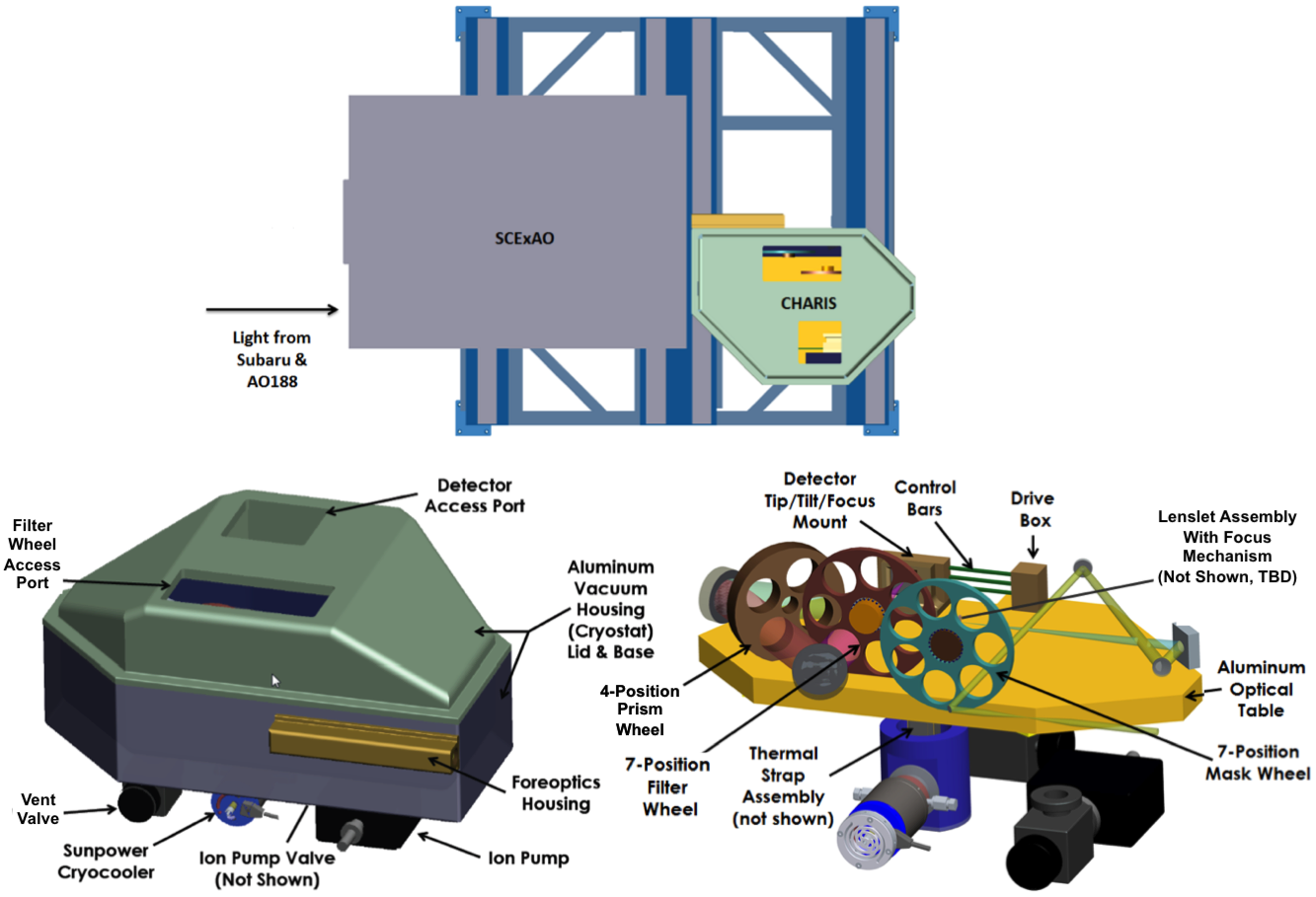}
\end{tabular}
\end{center}
\caption[example] 
{ \label{fig:MechPic1and2and3} 
 \emph{Top}: CHARIS mounted on the Nasmyth platform observation deck, shown from above.  \emph{Bottom left}: CHARIS vacuum housing.  \emph{Bottom right}: Optical layout on table.}
\end{figure}

CHARIS will be mounted on the Subaru telescope's Nasmyth platform observation deck as shown in Figure \ref{fig:MechPic1and2and3}, and will work in series with Subaru's existing AO188 and SCExAO instruments.  The CHARIS instrument is an aluminum vacuum cryostat enclosure (Figure \ref{fig:MechPic1and2and3}).  Mounted to the base of the cryostat is an ion pump to create high vacuum of the instrument down to $10^{-6}$ to $10^{-8}$ Torr vacuum and a Sunpower cryocooler to provide cryogenic cooling of ~15 W at 77 K.  Access ports on the cryostat will allow convenient access for maintenance and adjustments.  

The optical assembly is mounted on an aluminum plate with thermally isolated hard mounts.  A conceptual design of the optical component layout on the table is shown in Figure \ref{fig:MechPic1and2and3}.  Components mounted to the table are hard-mounted.  Components that require initial aligning will be accessible only with the top of the enclosure removed.  The detector will have active adjustment via remote motor control.  The Sunpower cryocooler (shown in Figure \ref{fig:MechPic1and2and3}) provides cooling via a silver strap that is attached to an OFHC bar located below the optical table.  This bar has silver straps along its length to distribute even cooling of the table, much like a tree trunk with lateral branches.

\begin{figure}
\begin{center}
\begin{tabular}{c}
\includegraphics[scale = 0.75]{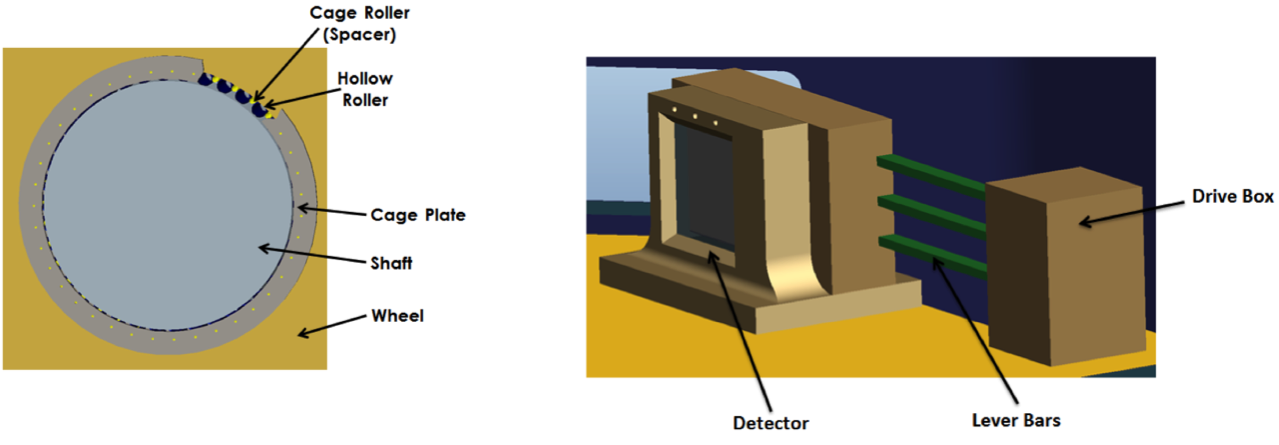}
\end{tabular}
\end{center}
\caption[example] 
{ \label{fig:Mech4and5} 
 \emph{Right}: Hollow roller bearing.  \emph{Left}: Detector tip/tilt/focus mount.}
\end{figure}

Three motorized wheel mechanisms will be used to position and reconfigure a set of filters, masks, and prisms in the optical layout.  Each of these wheels will rotate on a roller bearing that incorporates hollow rollers, as shown in Figure \ref{fig:Mech4and5}.  These rollers provide a spring-loaded preload engineered to provide adequate load carrying capacity.  They also provide a full line of contact to the inner and outer races (rather than the point contact provided by ball bearings), ensuring efficient heat transfer of the wheel assembly.  Efficient thermal coupling will ensure good thermal stability and shorter cool-down time.  

CHARIS will use a Teledyne Hawaii-2RG (H2RG) detector (Figure \ref{fig:Mech4and5}), housed in a tip/tilt/focus mechanism mount that uses a lever bar linkage system. The levers are push/pull fulcrum bars with a large ratio to provide $\sim1\mu m$ detector focus steps.  Each lever is actuated by a screw thermally isolated to a motor via a vacuum feedthrough.

%%%%%%%%%%%%%%%%%%%%%%%%%%%%%%%%%%%%%%%%%%%%%%%%%%%%%%%%%%%%%
\section{CONCLUSIONS} 
CHARIS will be the first exoplanet-purposed IFS on an 8m class telescope in the northern hemisphere able to achieve a small inner-working angle ($2 \lambda/D$) and high contrasts ($ 10^{-4}-10^{-7} $). CHARIS will provide R = 33 spectral resolution over a 1.75 arcsecond FOV. It is unique in that it will also offer both a low-resolution mode (R=14) able to collect imagery from $0.9-2.4 \mu m$ (y, J, H, and K bands) and a high resolution mode for R=65 spectra over a single bandpass (y, J, H or K band).  CHARIS will also implement innovative methods to minimize crosstalk to the $10^{-4}$ level. The instrument should achieve first light at the Subaru telescope by the end of 2015.

%%%%%%%%%%%%%%%%%%%%%%%%%%%%%%%%%%%%%%%%%%%%%%%%%%%%%%%%%%%%%
\acknowledgments       
 
This work was performed under a Grant-in-Aid for Scientific Research on Innovative Areas from MEXT of the Japanese government (Number 23103002).

%%%%%%%%%%%%%%%%%%%%%%%%%%%%%%%%%%%%%%%%%%%%%%%%%%%%%%%%%%%%%
%%%%% References %%%%%

%\bibliography{/Applications/TeX/bibLibrary/bibLibrary}

\bibliographystyle{spiebib}

\end{document}